# The role of silanol nests in the activation of the $[Fe=O]^{2+}$ group in the reaction of hydrogen atom transfer from methane.

V.Yu. Kovalskii[1]

[1]Boreskov Institute of Catalysis SB RAS, Lavrentieva ave. 5, Novosibirsk, Russia, 630090

## Abstract

Silanol nests can play the role of places into which positively charged groups, such as, $[FeO]^{2+}$, can invade. In the framework of this work, the influence of such structures on the activity of the [FeO]2+ group in the reaction of detachment of a hydrogen atom from methane was considered. Two ways of the reaction of hydrogen atom transfer (HAT) from methane were found: the so-called ferryl and oxyl routes. It was shown that the reaction of the detachment of a hydrogen atom from methane, which is a limiting stage of the oxidation of methane to methanol at the alpha center, proceeds through the formation of the so-called oxidation state [Fe(III)-O(-)]2+, and fingerprint of this state is negative spin density on oxo moiety.

## Introduction

The methane binding problem is one of the most difficult problems, since methane oxidation requires breaking or at least weakening the C-H bond, whose energy is about 140 kcal/mol. Some living organisms are able to absorb methane from the gas phase due to the presence of enzymes that can bind methane under mild conditions, such as methane monooxygenase. The complexity of using enzymes in industry is that it is difficult to organize large-tonnage production. It is known that a large number of these enzymes contains the group [M=O] with active oxygen, on which the partial oxidation of methane takes place. Based on information on the structure of these enzymes, a large number of biomimetic catalysts were synthesized and tested, but they are all inferior in oxidizing ability to enzymes and are also difficult to use in large-tonnage production, as are enzymes.

To solve this problem, there were attempts to create systems containing [M = O] based on, among other things, zeolites. One of the successful examples of this approach can be considered the so-called Panov alpha-oxygen, capable of oxidizing methane and benzene under mild conditions with a fairly high [1,2]. Despite a fairly lengthy study of this system, the structure of the active center has not yet been uniquely determined. The active center model in the form of the monomer [Fe = O] 2+ with a bond length of about 1.6 Å localized in the cation exchange position of the zeolite lattice predominates.[3] In some works, it is represented as an iron dimer, but only one oxygen atom acts as an active fragment. [4] The literature also mentions the possibility of the formation of iron trimers in the structure of Fe-ZSM-5, similar to the structure of the active center of ferredoxin II, Fe3S4.[5] In the literature, most often the group containing iron is placed in a six-membered ring, as the most likely place for the introduction of such a structure. In addition to such places, there are so-called "hydroxyl nests" (hydroxyl or silanol nests). "Hydroxyl nests" are formed during dealumination of the initial zeolite, as a result of which tetrahedral aluminum is removed from the zeolite framework.

So, for example, when processing in solutions of inorganic acids at a pH of less than 4, the structure of zeolite Y is destroyed, while at a pH of more than 4, partial amorphization and simultaneous dealumination of the zeolite framework occurs. In this case, 2 reactions proceed sequentially: decationation and dealumination due to the formation of the Brønsted acid center (BCC).

Aluminum in the BCC is weaker bound to the framework and leaves it.[6] In this case, a structural defect is formed - the "hydroxyl nest", consisting of four hydroxyl groups. (Figure 1)

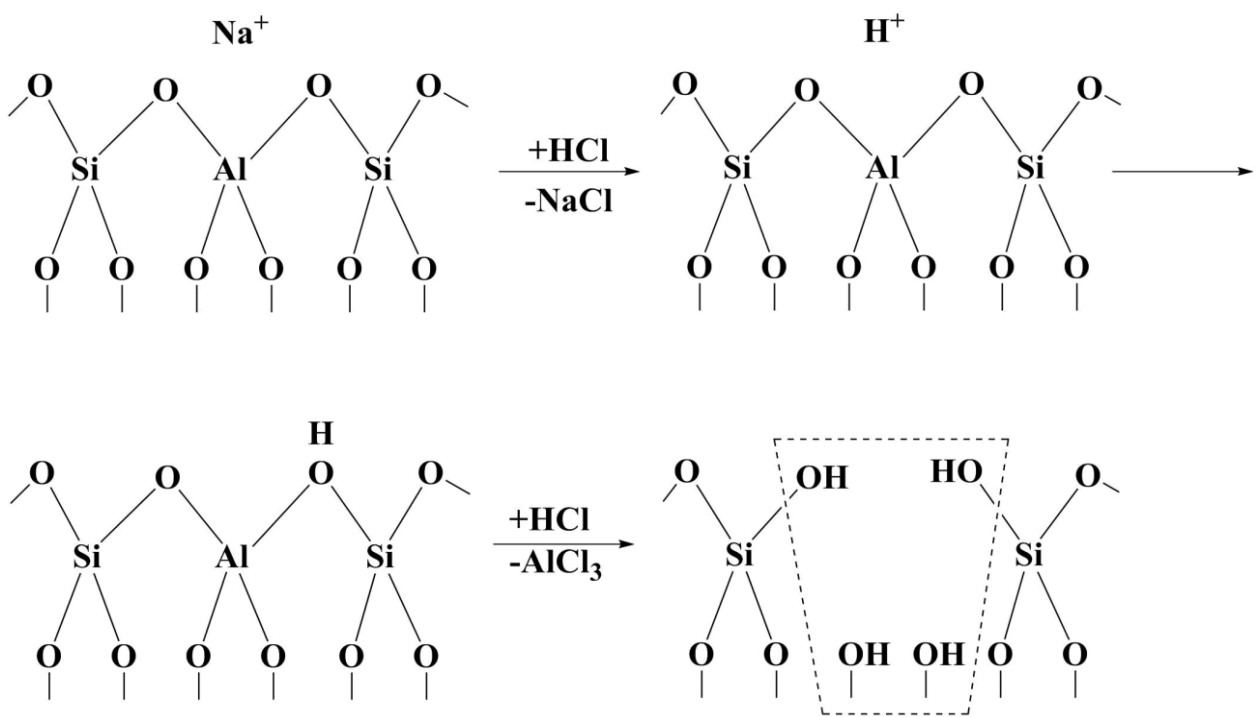

*Figure 1. The scheme of formation of the "hydroxyl nest", consisting of four hydroxyl groups.*

"Hydroxyl nests" can play the role of places into which positively charged groups, such as, $[FeO]^{2+}$, can invade. In the framework of this work, the influence of such structures on the activity of the $[FeO]^{2+}$ group in the reaction of detachment of a hydrogen atom from methane was considered. Earlier in our works, using the example $O=Fe(OH)_2$, two ways of the reaction of the detachment of a hydrogen atom from methane were found: the so-called ferryl (generally, amyl) and oxyl routes. It was shown that the reaction of the detachment of a hydrogen atom from methane, which is a limiting stage of the oxidation of methane to methanol at the alpha center, proceeds through the formation of the so-called oxidation state $[Fe---O^*]^{2+}$, and fingerprint of this state is negative spin density on oxo moiety.[7] It should be noted that in many studies, the oxidized state refers to the Fe–O group with an elongated bond (up to 1.7–1.8 Å) and an oxygen center with some positive spin density, which is interpreted as evidence of the radical nature of this center. [8] However, in specific calculations, both in the ferryl and in the oxyl state, there is a noticeable magnitude of spin density, positive and negative signs, respectively.[?] A negative spin density oxo center was first identified by Morokuma [9]. Morokuma et al., When considering the reaction of the detachment of a hydrogen atom from methane in a methane monoxygenase model, found that in the transition state a negative spin density appears on the terminal oxygen atom and a methyl radical with a negative spin density forms. The authors associated the appearance of negative spin density on the methyl radical with the fact that two iron atoms are ferromagnetically coupled; therefore, the methyl radical must bind antiferromagnetically in order to satisfy the total spin moment of 9.

Baerends investigated the active center models of the alpha center containing $[FeO]^{2+}$ in the form $O=M (H_2O)_5^{2+}$ by DFT [10] and by Car-Parinello method [11]. To explain the appearance of negative spin density on the methyl radical, the authors proposed a mechanism involving the transfer of α electron from methane to σ* the iron orbital. Thus, five unpaired α electrons are formed on the iron atom, and the β electron remains on the methyl radical. Baerends et al suggested that the activity of the $[M=O]^{2+}$ group correlates with the position of the first unoccupied orbital

(σ*). The lower the energy of this orbital, the higher the activity, estimated by the magnitude of the hydrogen separation barrier.

But this mechanism has several disadvantages. So, for example, he cannot explain the oxidative nature of the transition state, which is noted in many computational works [8,12] in which the unoccupied orbital σ* is filled with its own electron and becomes inaccessible to the hydrogen electron. But this mechanism does not explain the appearance of negative spin density on oxygen during the reaction in the early stages.

Nesse in his work [8] proposes such a mechanism: oxidative oxygen is more active than oxoligand and has a higher electrophilicity. Thus, it is possible to imagine the activation of the C – H bond as a process consisting of a preparatory stage, at which the curve intersects between the ground state of $[FeO]^{2+}$ and the state of charge transfer from the ligand to the metal, followed by the stage of detachment of the hydrogen atom. As explained by PCET theory [doi:10.1146/annurev.physchem.49.1.337], the latter can proceed as proton transfer followed by electron transfer, vice versa or simultaneously. The activity of $[FeO]^{2+}$ Neese associates with the fact that, at the preparatory stage, the ability of $[FeO]^{2+}$ to interact with the CH binding orbital is enhanced, since the O-$p_z$ orbital overlaps with the CH σ bond more efficiently than the Fe- $d_{z2}$ antibonding orbital, which includes only the limited nature of O-$p_z$. Enhanced overlap greatly facilitates electron transfer and the formation of three-center MOs observed in the transition state. The oxyl-oxygen is intrinsically much more reactive than the oxo-ligand and is highly electrophilic. Thus, one may picture the C–H bond activation process as consisting of a preparatory step in which there is a curve crossing between the ground state of the $(FeO)^{2+}$ core and a LMCT (Ligand-to-metal (ion) charge transfer) state, followed by the genuine C–H abstraction step. The preparatory step greatly enhances the ability of the $(FeO)^{2+}$ core to interact with the bonding C–H s-orbital since the O-$p_z$ orbital more efficiently overlaps with the C–Hs-bond than the Fe-$d_z$2 antibonding orbital that only involves limited O-$p_z$ character. The enhanced overlap greatly facilitates electron transfer and formation of the three-center MOs observed in transition state.

Solomon: [13]

The mechanism for activating the iron–oxo intermediate for electrophilic reactivity is the same for both, aromatic electrophilic attack and hydrogen-atom abstraction: Elongation of the Fe-O bond transforms the ferryl-oxo, $Fe^{IV}=O^{2-}$, to a ferric-oxyl, $Fe^{III}$-$O^{•-}$ species in the transition state. The short Fe-O bond in the $Fe^{IV}$=O intermediate allows for large σ- and especially π-overlap between the iron d- and oxo p-orbitals, forming a strong covalent bond, which enables significant charge donation from the oxo to the iron to stabilize the high oxidation state. Elongation of the Fe-O bond decreases the overlap and significantly reduces the charge donation. In this case, one electron is transferred from the oxo to the iron, which results in significant spin polarization and the bi-radical character of the ferric-oxyl transition state. The dominant radical character on the oxygen significantly increases its electrophilicity, enhancing the reactivity with the electron density on the aromatic π-system or the C-H σ-bond.

In the transition state for hydrogen-atom abstraction, energy is required to partially break the C-H bond. Thus, this activation energy is somewhat dependent on the C-H bond strength of the substrate.

Earlier, we studied systems containing the $[Fe=O]^{2+}$: monomer O=Fe(OH)$_2$, dimer OFe$_2$(OH)$_5$ and tetramer OFe$_4$(μ-O)$_4$(OH)$_3$. When simulating the first stage of C-H bond C-H:

[Fe=O]$^{2+}$+CH$_4$→[FeOH]$^{2+}$+CH$_3$• two reaction routes, the so-called oxyl and ferryl routes, were detected.

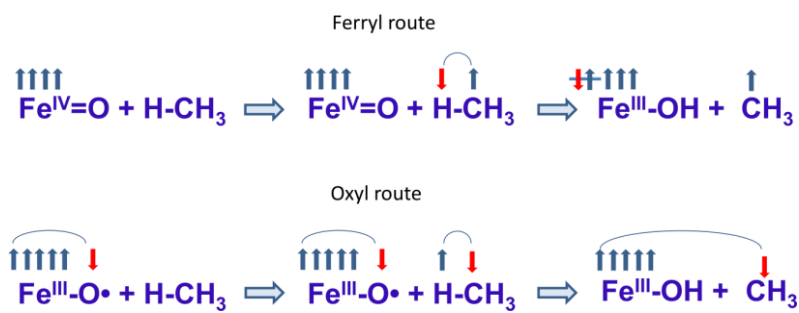

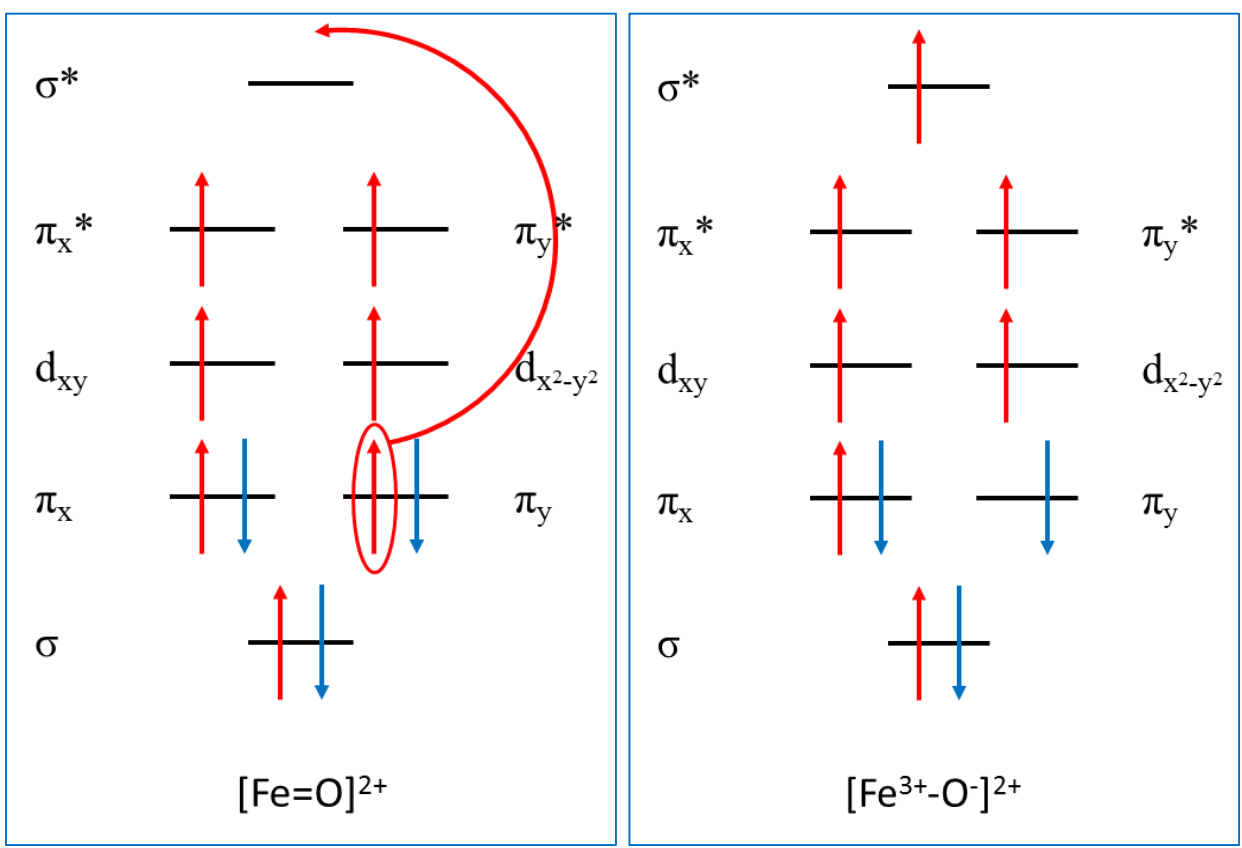

*Figure 2. Scheme of formation oxyl state [Fe$^{3+}$-O$^-$]$^{2+}$ from ferryl state [Fe=O]$^{2+}$.*

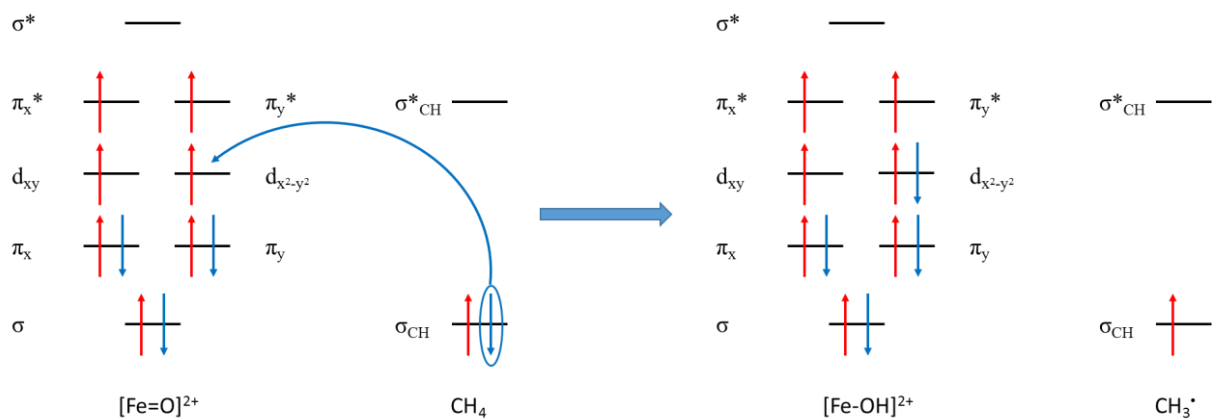

*Figure 3. Ferryl route [FeO]$^{2+}$ +CH$_4$.*

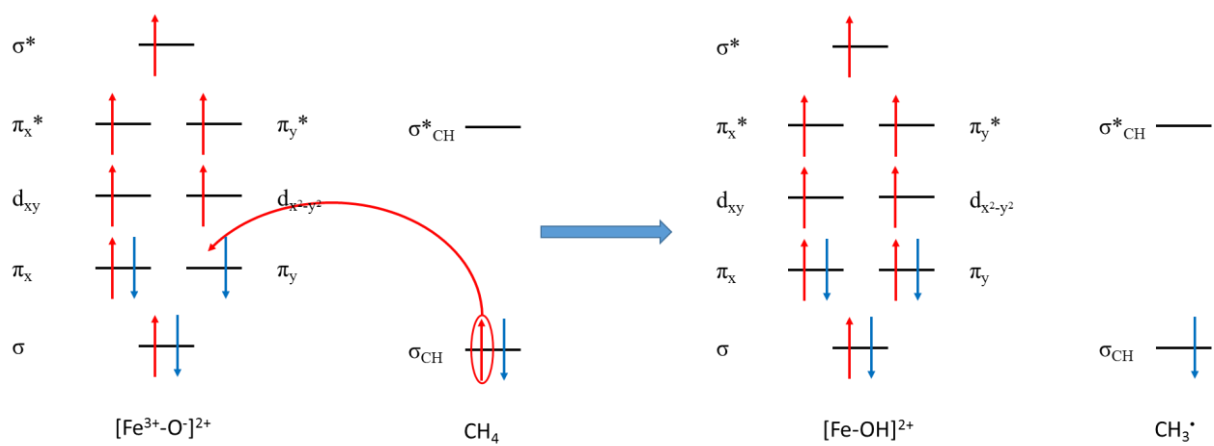

*Figure 4. Oxyl route [FeO]$^{2+}$ + CH$_4$.*

In reality, the reaction from the hydrogen atom proceeds through the oxyl state Figure 2. Then the reaction proceeds in the oxide stream in accordance with the instructions, Figure 4. The ferryl path can be realized, but the barrier reaction is due to higher than in the oxyl scenario. [14]

In this paper, we solve the problem of studying the possible effect of structures on the similarity of "hydroxyl nests" on the state and reactivity of the oxo center of the ferryl group in simple models containing Al and Ga atoms. Various model structures of the Fe-ZSM-5 cluster from simple to more complex are considered: O=Fe(OM(OH)$_2$)$_2$, O=Fe(M(OH)$_2$)$_2$($\mu$-O)$_4$(Si(OH)$_2$)$_2$ и 6-ring model ZSM-5 (FeM2Si5H12O21). The first one (O=Fe(OM(OH)$_2$)$_2$) were formed by changing hydrogen atom at O=Fe(OH)$_2$ by M(OH)$_2^+$, where M=Al, Ga. Thus, we can try to take into account the influence of the ligand environment in a rather simple approximation. The second model, O=Fe(M(OH)$_2$)$_2$($\mu$-O)$_4$(Si(OH)$_2$)$_2$, was formed by forming a cation-exchange position in a four-membered silicon cycle. The latter model was formed on the basis of the model of the active center FeZSM-5, in which the model is simplified to the maximum to facilitate the calculation, but at the same time retaining the active zeolite ring. [15].

## ZSM-5 Models

O=Fe(OM(OH)2), M=Al, Ga.

To determine the effect of the ligand environment on the [FeO] 2+ activity, the previously studied model particle of α-oxygen O = Fe (OH) 2 was taken and the hydrogen atom in it was replaced by the group M (OH) 2, where M = Al, Ga. This group has the same symmetry as O = Fe (OH) 2;

thus, it is possible to evaluate the effect of the ligand environment on the stabilization of the oxide state and the change in the separation barrier of the hydrogen atom from methane.

Al

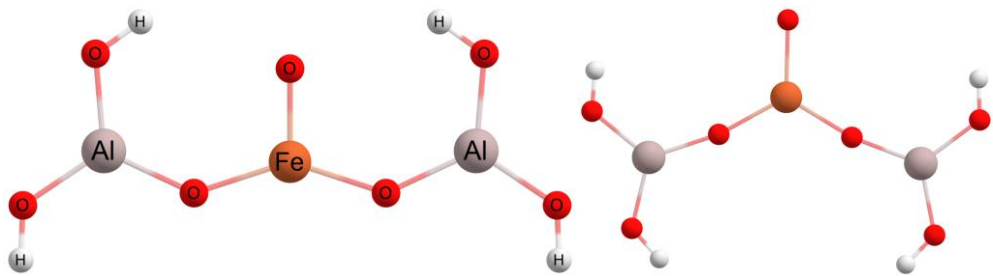

Figure 5. The O=Fe(OAl(OH)$_2$) model: $^5A_1$ and $^5B_1$ states.

In a previously published article [16] for the O=Fe(OH)$_2$ system the authors showed that the $^5B_1$ и $^5B_2$ oxyl state is higher in energy relative to the $^5A_1$ ground state by 42 and 21 kcal/mol, respectively, and $^5A_2$ – by 51 kcal/mol. For our model, $^5A_2$ is also by 53 kcal/mol, and the $^5B_1$ и $^5B_2$ oxyl states are 20 и 10.5 kcal/mol higher, which is significantly lower than for O=Fe(OH)$_2$ (Table 1). That is, even in such a simple model, the influence of the ligand environment on the stabilization of the oxide state is significant.

CH$_4$ oxidation

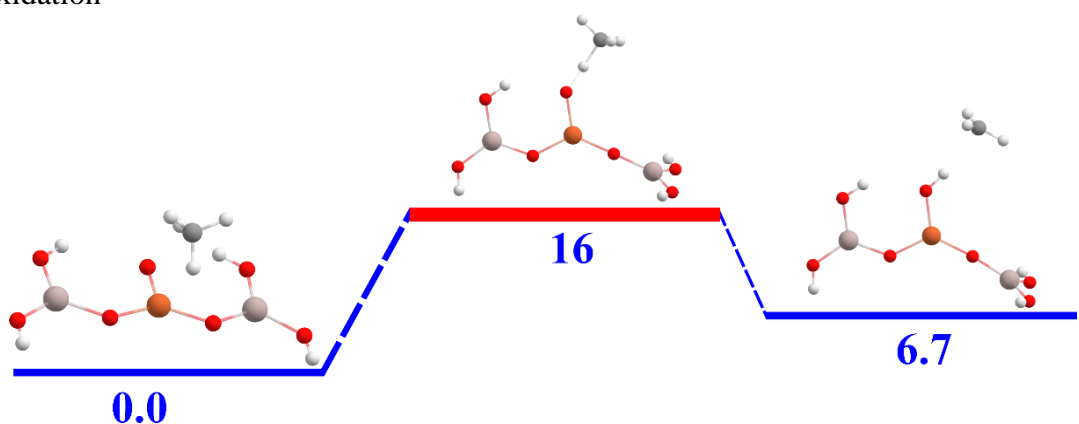

Figure 6. Reaction path O=Fe(OAl(OH)$_2$)+CH$_4$. Energies in kcal/mol relative to reactants.

Despite the stabilization of the oxide state, the hydrogen atom detachment barrier, 16 kcal/mol (Table 2), does not differ much from the previously published value for the oxyl – 18 kcal/mol [14]. The reaction proceeds through an oxide transition state, the hallmark of which is the emerging negative spin density on the methyl, $q_s(C) = -0.490$ (Table 2). The angle ∠(O-Fe-C) is about 130 degrees, which corresponds to a π-attack, as previously indicated in an article by Shaik et al. [17] The authors argue that at $S_z=2$ the angle of attack of approximately 120° should correspond to the formation of a methyl radical with α-spin, that is, with a positive spin density. In our calculations, when the angle of attack is 130°, which corresponds to a π-attack, a negative spin density is formed on the methyl radical in the intermediate. Therefore, we suggested that the mechanism proposed by Shaik is not correct.

Ga

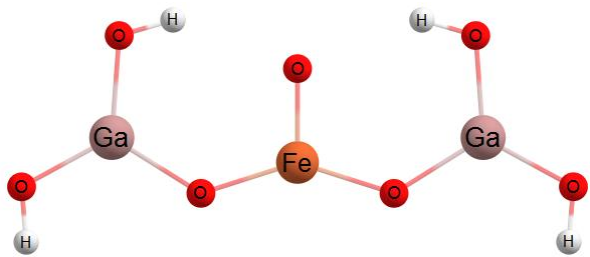

Figure 7. The O=Fe(OGa(OH)$_2$) model: NoSymm state.

CH$_4$ oxidation

When Al is replaced by Ga, the separation barrier of the hydrogen atom from methane by O=Fe(OGa(OH)$_2$) practically doesn't change and is equal to 18 kcal/mol. Thus, we can conclude that the direct influence of the metal on the course of the reaction is practically absent.

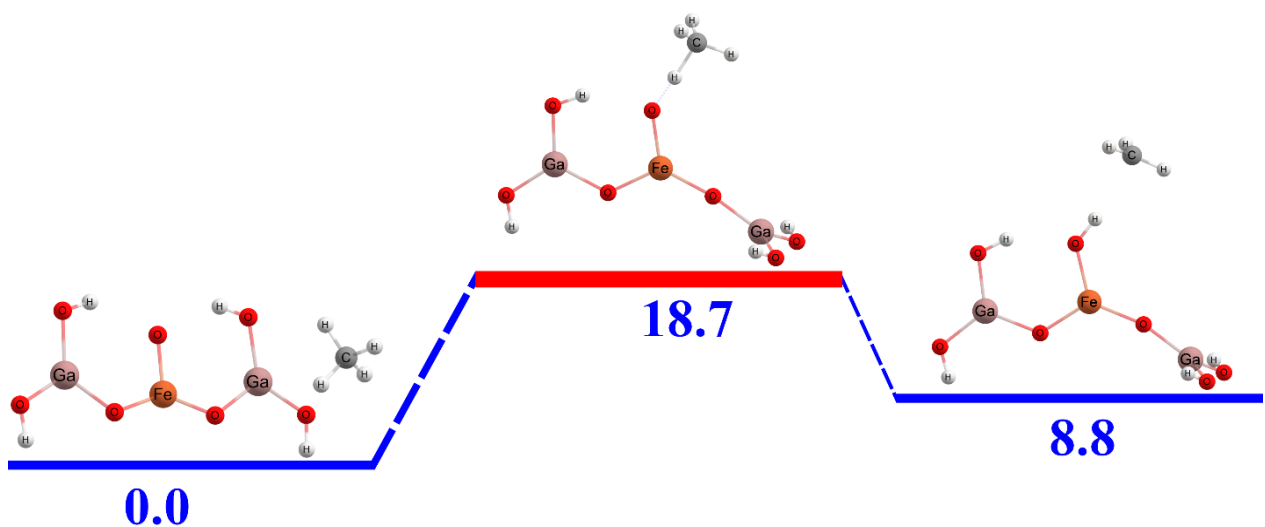

*Figure 8. . Reaction path O=Fe(OGa(OH)$_2$)+CH$_4$. Energies in kcal/mol relative to reactants.*

O=Fe(M(OH)$_2$)$_2$(μ-O)$_4$(Si(OH)$_2$)$_2$, M=Al, Ga.

Al

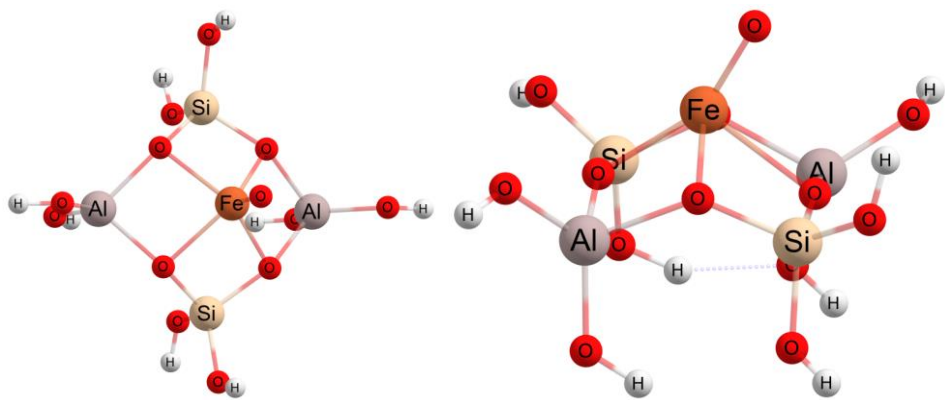

Figure 9. O=Fe(M(OH)$_2$)$_2$(μ-O)$_4$(Si(OH)$_2$)$_2$ Model: $^5A_1$ and NoSymm structure.

When using the four-membered cycle, it is possible to significantly stabilize the oxyl states ($^5B_1$ and $^5B_2$); they are comparable in energy to $^5A_1$ (Table 5). When symmetry is disabled (NoSymm structure), the energy of the system drops by 9.4 kcal / mol (Table 5). Therefore, we can assume that a system with artificially specified symmetry is a stressed structure. Under certain circumstances, if the environment can create such a tense structure, the oxidation state can become the ground state. One of such circumstances may be a geometric factor, and specifically ∠(O-Fe-O), as it was previously shown that the influence of this angle on the stabilization of the oxidation state is significant. [18] So for the oxyl state is characterized by lower values of the angle ∠(O-Fe-O).

CH$_4$ oxidation

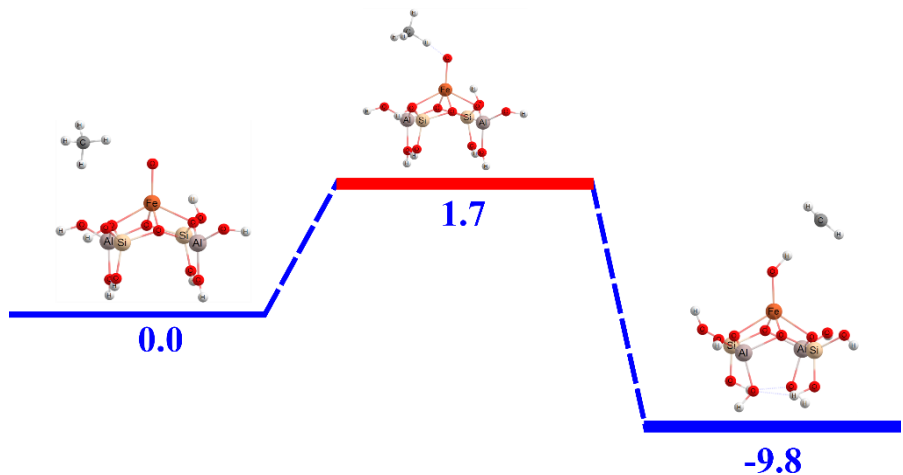

*Figure 10. Reaction path O=Fe(Al(OH)$_2$)$_2$(μ-O)$_4$(Si(OH)$_2$)$_2$+CH$_4$. Energies in kcal/mol relative to reactants.*

For a structure with a four-membered ring, the hydrogen atom detachment reaction barrier is only 1.7 kcal/mol. Therefore, for structures like these, a very high activity in the HAT reaction can be expected. Такие структуры могут образоваться в процессах формирования ранее упомянутых «гидроксильных гнёзд».

Ga

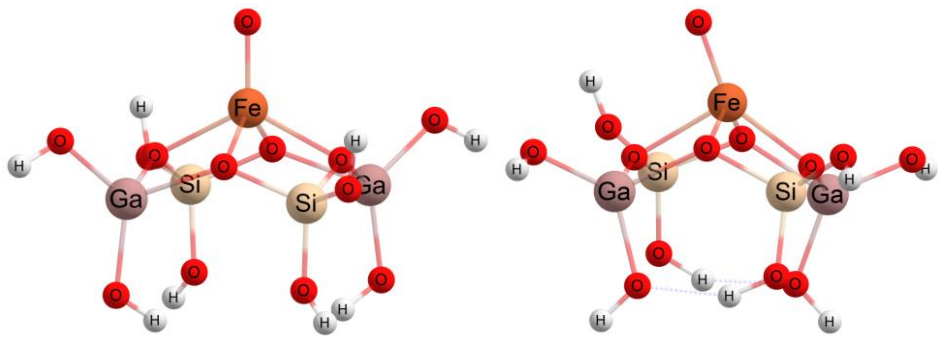

Figure 11. The O=Fe(M(OH)2)2(μ-O)4(Si(OH)2)2 model: $^5A_1$ and NoSymm structure.

CH₄ oxidation

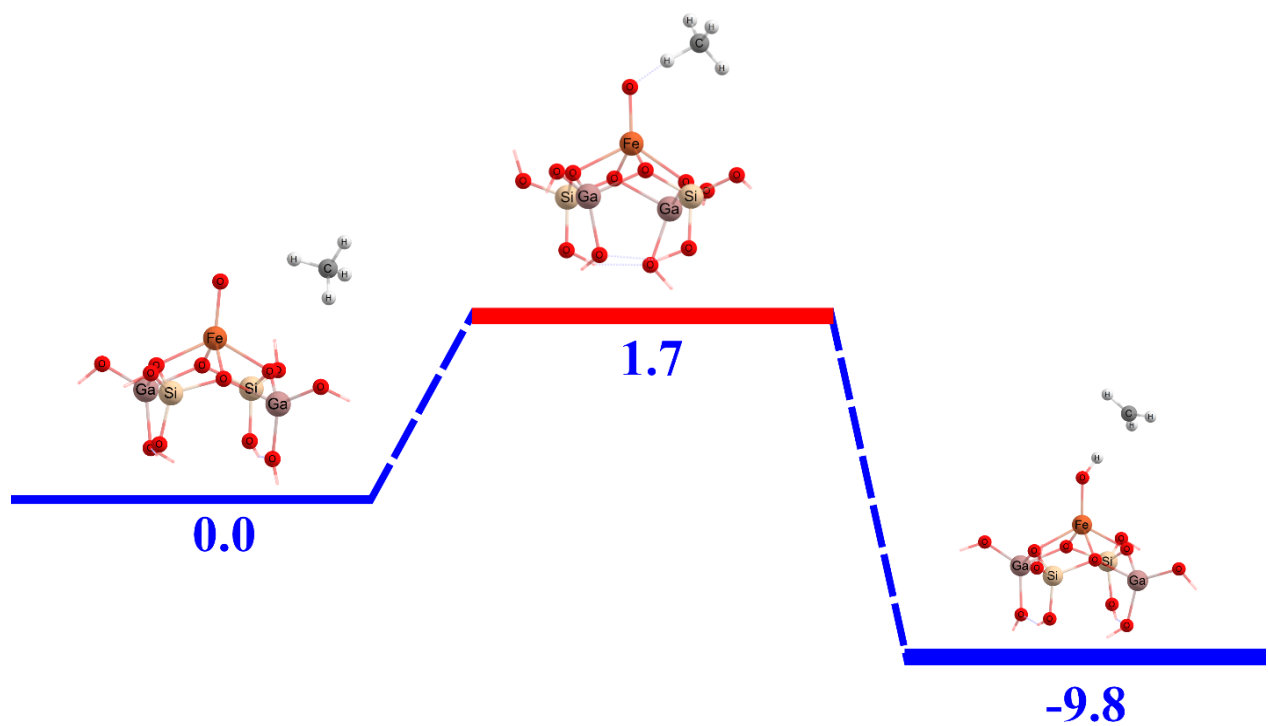

Figure 12. Reaction path O=Fe(Ga(OH)₂)₂(μ-O)₄(Si(OH)₂)₂+CH₄. Energies in kcal/mol relative to reactants.

6-ring model ZSM-5 (FeM2Si5H12O21, M=Al, Ga.)
Al

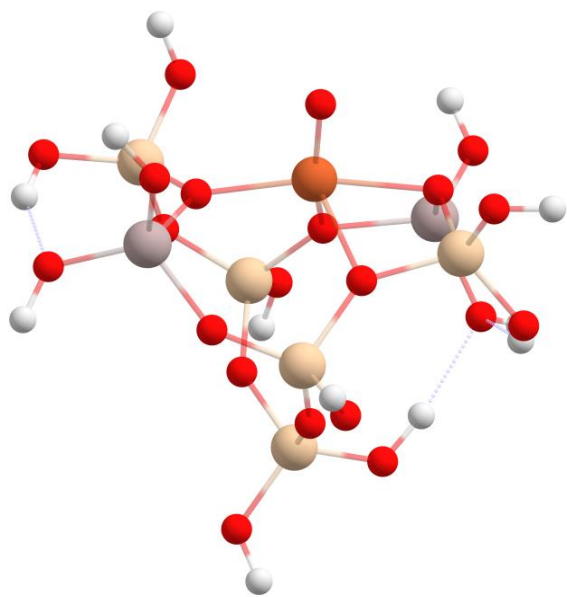

Figure 13. 6-ring model ZSM-5 (FeAl2Si5H12O21)

CH$_4$ oxidation

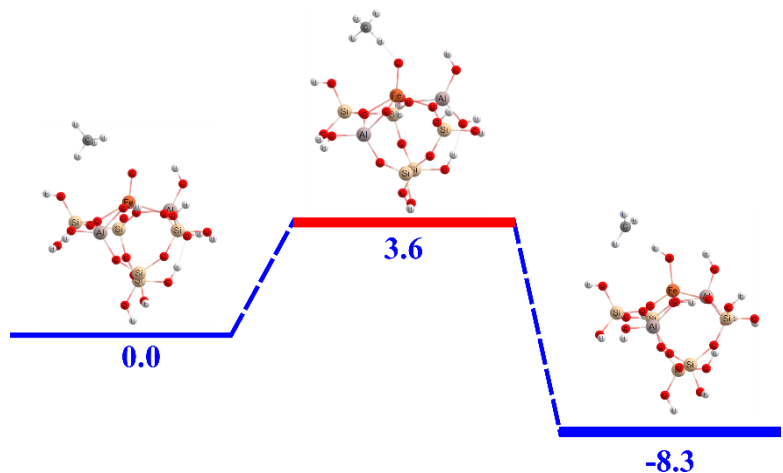

*Figure 14. Reaction path FeAl2Si5H12O21+CH$_4$. Energies in kcal/mol relative to reactants.*

Ga

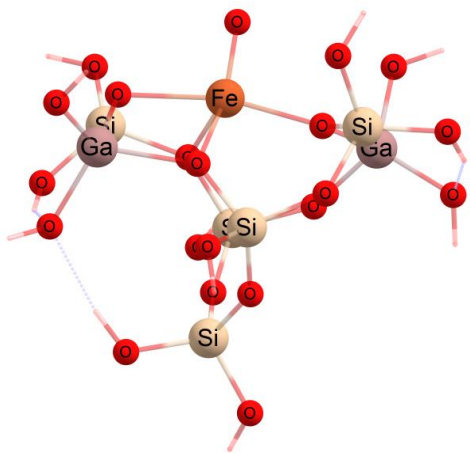

Figure 15. 6-ring model ZSM-5 (FeGa2Si5H12O21)

CH$_4$ oxidation

HAT on 6-ring model протекает с барьером, сопоставимым с ранее приведённым в литературе барьером для модели цеолита Fe-ZSM-5 [15] и для модели гидроксида железа в виде кубана O=Fe$_4$(μ-O)$_4$(OH)$_3$ [7].

# Discussion

# Acknowledgments

The reported study was funded by RFBR according to the research project № 18-33-00932.

# Supporting materials

# Computational details

All calculations were performed using by Gaussian'09 package [19] on UB3LYP/6-311++G(d,p) level. [20,21] All systems were uncharged and in quintet state ($S_z=2$).

SCF=(tight,Conver=8) int=(grid=UltraFine)

Charge = 0 Multiplicity = 5.

*Table 1. The different states of O=Fe(OAl(OH)$_2$)$_2$ model in C2v point group: energy, geometrical parameters, spin densities, <$S^2$> and energy of 1s-orbital of Fe and O from group [FeO]$^{2+}$.*

| Symmetry | $^5A_1$ | $^5A_2$ | $^5B_1$ | $^5B_2$ | NoSymm |
|----------|---------|---------|---------|---------|--------|

| Energy, a.u. | -2278.12981301 | -2278.04541890 | -2278.09727860 | -2278.11304619 | -2278.12981432 |
|---|---|---|---|---|---|
| E$_{rel}$, kcal/mol | 0 | 52.95892781 | 20.41647339 | 10.52216087 | 0 |
| d(Fe=O), Å | 1.62553 | 1.73739 | 1.80953 | 1.71269 | 1.62562 |
| ∠(O-Fe=O) | 108.124 | 110.540 | 120.278 | 122.753 | 108.115 |
| ∠(O-Fe-O) | 143.752 | 138.920 | 119.444 | 114.493 | 143.768 |
| <S$^2$> | 6.0596 | 6.5835 | 6.8115 | 6.5558 | 6.0596 |
| q$_s$(Fe) | 3.176 | 3.641 | 4.133 | 3.897 | 3.176 |
| q$_s$(O) | 0.544 | 0.983 | -0.688 | -0.399 | 0.544 |
| E(Fe1s), a.u. | -256.193568 | -256.161139 | -256.197110 | -256.195859 | -256.193620 |
| E(O1s), a.u. | -19.172867 | -19.129572 | -19.166425 | -19.173770 | -19.172910 |
| E(Fe1s), eV | -6971,382075 | -6970,499637 | -6971,478457 | -6971,444416 | -6971,38349 |
| E(O1s), eV | -521,720285 | -520,5421681 | -521,5449893 | -521,7448569 | -521,7214551 |
| E$_{rel}$(Fe1s), eV | 0 | 0,882438 | -0,096382 | -0,062341 | -0,001415 |
| E$_{rel}$(O1s), eV | 0 | 1,1781169 | 0,1752957 | -0,0245719 | -0,0011701 |

E$_{ads}$ = E(NoSymm+CH$_4$) - (E(NoSymm) + E(CH$_4$)) = -2318.66454894-(-2278.12981432-40.5339575449) = -0,000777075 (a.u.) = -0,487622007 (kcal/mol)

*Table 2. Characteristics of adsorbed system, reactants, transition state and products for O=Fe(OAl(OH)$_2$)+CH$_4$.*

| Structure | NoSymm+CH$_4$ | Reactants | TS | Products |
|---|---|---|---|---|
|  | 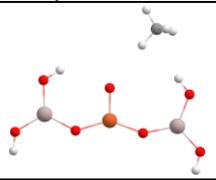 | 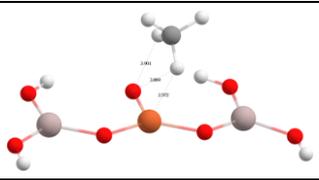 | 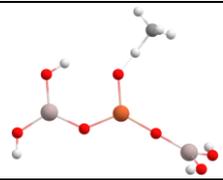 | 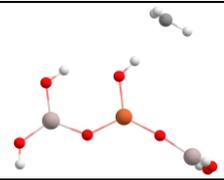 |
| Energy, a.u. | -2318.66454894 | -2318.66454600 | -2318.63879039 | -2318.65391320 |
| E$_{rel}$, kcal/mol | 0 | 0,001844878 | 16,16373483 | 6,67402789 |
| d(Fe=O), Å | 1.62442 | 1.62676 | 1.75919 | 1.82181 |
| ∠(O-Fe=O) | 106.763 | 108.071 | 108.858 | 110.151 |
| ∠(O-Fe-O) | 142.431 | 143.376 | 127.478 | 125.614 |
| ∠(Fe-O-H) | 143.24 | 72.025 | 137.805 | 128.115 |
| ∠(Fe-O-C) | 141.660 | 62.068 | 138.431 | 134.307 |
| <S$^2$> | 6.0607 | 6.0601 | 6.7128 | 7.0050 |
| q$_s$(Fe) | 3.184 | 3.180 | 3.999 | 4.145 |
| q$_s$(O) | 0.540 | 0.539 | -0.034 | 0.294 |
| q$_s$(H) | 0 | 0.001 | 0.015 | -0.029 |
| q$_s$(C) | -0.001 | 0.006 | -0.490 | -1.075 |
| E(Fe1s), a.u. | -256.192350 | -256.191130 | -256.176520 | -256.176540 |
| E(O1s), a.u. | -19.171040 | -19.170210 | -19.147460 | -19.135720 |
| E(Fe1s), eV | -6971,348931 | -6971,315733 | -6970,918175 | -6970,918719 |
| E(O1s), eV | -521,6705698 | -521,6479844 | -521,0289254 | -520,7094637 |
| E$_{rel}$(Fe1s), eV | 0 | 0,033198 | 0,430756 | 0,430212 |
| E$_{rel}$(O1s), eV | 0 | 0,0225854 | 0,6416444 | 0,9611061 |

*Table 3. The different states of O=Fe(OGa(OH)$_2$)$_2$ model in C2v point group: energy, geometrical parameters, spin densities, <S$^2$> and energy of 1s-orbital of Fe and O from group [FeO]$^{2+}$.*

| Symmetry | $^5$A$_1$ | $^5$A$_2$ | $^5$B$_1$ | $^5$B$_2$ | NoSymm |
|---|---|---|---|---|---|
| Energy, a.u. | -5642.78586202 |  |  |  | -5642.78586431 |
| E$_{rel}$, kcal/mol |  |  |  |  |  |
| d(Fe=O), Å | 1.63225 |  |  |  | 1.63220 |
| ∠(O-Fe=O) | 107.805 |  |  |  | 107.792 |
| ∠(O-Fe-O) | 144.389 |  |  |  | 144.410 |
| <S$^2$> | 6.0583 |  |  |  | 6.0583 |
| q$_s$(Fe) | 3.159 |  |  |  | 3.159 |
| q$_s$(O) | 0.533 |  |  |  | 0.533 |
| E(Fe1s), a.u. | -256.196426 |  |  |  | -256.196439 |
| E(O1s), a.u. | -19.175179 |  |  |  | -19.175195 |
| E(Fe1s), eV. | -6971,459845 |  |  |  | -6971,460199 |

| E(O1s), eV | -521,7831978 | | | -521,7836332 |
| E$_{rel}$(Fe1s), eV. | 0 | | | -0,000354 |
| E$_{rel}$(O1s), eV | 0 | | | -0,0004354 |

E$_{ads}$ = E(NoSymm+CH$_4$) - (E(NoSymm) + E(CH$_4$)) = -5683.32060760-(-5642.78586431-40.5339575449) = -0.000785745 (a.u.) = -0.493062515 (kcal/mol)

*Table 4. Characteristics of adsorbed system, reactants, transition state and products for O=Fe(OGa(OH)$_2$)+CH$_4$*

| Structure | NoSymm+CH$_4$ | Reactants | TS | Products |
|---|---|---|---|---|
| | 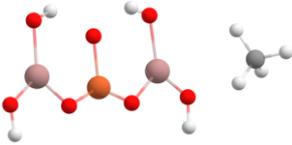 | 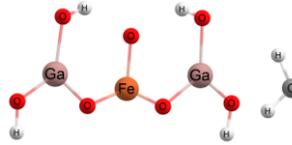 | 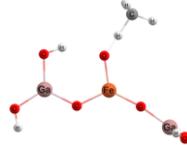 | 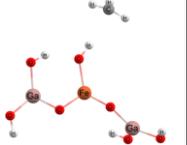 |
| Energy, a.u. | -5683.32060760 | -5683.32060472 | -5683.29165928 | -5683.30658086 |
| E$_{rel}$, kcal/mol | 0 | 0 | 18,17 | 8,8 |
| d(Fe=O), Å | 1.63237 | 1.63237 | 1.76131 | 1.82619 |
| ∠(O-Fe=O) | 107.806 | 107.806 | 108.452 / 124.864 | 109.631 / 124.947 |
| ∠(O-Fe-O) | 144.368 | 144.374 | 126.683 | 125.422 |
| ∠(Fe-O-H) | 87.121 | 87.051 | 137.973 | 128.488 |
| ∠(Fe-O-C) | 76.586 | 76.505 | 138.786 | 134.257 |
| <S$^2$> | 6.0583 | 6.0583 | 6.7086 | 7.0056 |
| q$_s$(Fe) | 3.158 | 3.158 | 3.980 | 4.118 |
| q$_s$(O) | 0.533 | 0.533 | -0.046 | 0.276 |
| q$_s$(H) | 0 | 0 | -0.002 | -0.038 |
| q$_s$(C) | 0.001 | 0.001 | -0.484 | -1.071 |
| E(Fe1s), a.u. | -256.195370 | -256.195370 | -256.177040 | -256.176870 |
| E(O1s), a.u. | -19.174100 | -19.174100 | -19.149210 | -19.138290 |
| E(Fe1s), eV. | -6971,43111 | -6971,43111 | -6970,932325 | -6970,927699 |
| E(O1s), eV | -521,7538367 | -521,7538367 | -521,0765453 | -520,779397 |
| E$_{rel}$(Fe1s), eV. | 0 | 0 | 0,498785 | 0,503411 |
| E$_{rel}$(O1s), eV | 0 | 0 | 0,6772914 | 0,9744397 |

*Table 5. The different states of O=Fe(Al(OH)$_2$)$_2$(µ-O)$_4$(Si(OH)$_2$)$_2$ model in C2v point group: energy, geometrical parameters, spin densities, <S2> and energy of 1s-orbital of Fe and O from group [FeO]$^{2+}$.*

| Symmetry | $^5A_1$ | $^5A_2$ | $^5B_1$ | $^5B_2$ | NoSymm |
|---|---|---|---|---|---|
| Energy, a.u. | -3311.54420701 | -3311.50143231 | -3311.54568887 | -3311.54706312 | -3311.55915096 |
| E$_{rel}$, kcal/mol | 0 | 26,84153061 | -0,929881228 | -1,792236158 | -9,377470593 |
| d(Fe=O), Å | 1.60328 | 1.70177 | 1.75535 | 1.75709 | 1.60708 |
| ∠(O-Fe=O) | 118.090 | 115.301 | 119.950 | 119.976 | 108.176 / 98.418 |
| ∠(O-Fe-O) | 123.820 / 123.820 | 129.397 / 129.397 | 120.101 / 120.101 | 120.047 / 120.047 | 134.909 / 107.991 |
| <S$^2$> | 6.1779 | 6.1013 | 6.7496 | 6.7531 | 6.0774 |
| q$_s$(Fe) | 3.303 | 2.883 | 4.012 | 4.044 | 3.148 |
| q$_s$(O) | 0.258 | 1.246 | -0.552 | -0.566 | 0.565 |
| E(Fe1s), a.u. | -256.215664 | -256.170853 | -256.220635 | -256.220401 | -256.211729 |
| E(O1s), a.u. | -19.201990 | -19.141748 | -19.195135 | -19.195061 | -19.193293 |
| E(Fe1s), eV. | -6971,983337 | -6970,763968 | -6972,118605 | -6972,112238 | -6971,876261 |
| E(O1s), eV | -522,5127622 | -520,8734939 | -522,3262282 | -522,3242145 | -522,2761048 |
| $E_{rel}$(Fe1s), eV. | 0 | 1,219369 | -0,135268 | -0,128901 | 0,107076 |
| $E_{rel}$(O1s), eV | 0 | 1,6392683 | 0,186534 | 0,1885477 | 0,2366574 |

$E_{ads}$ = E(NoSymm+$CH_4$) - (E(NoSymm) + E($CH_4$)) = -3352.09365313-(-3311.55915096-40.5339575449) = -0.000544625 (a.u.) = -0.341757424 (kcal/mol)

*Table 6. Characteristics of adsorbed system, reactants, transition state and products for $O=Fe(Al(OH)_2)_2(\mu-O)_4(Si(OH)_2)_2 +CH_4$*

| Structure | NoSymm+$CH_4$ | Reactants | TS | Products |
|---|---|---|---|---|
| Energy, a.u. | -3352.09365313 | -3352.10086859 | -3352.09823616 | -3352.11650537 |
| $E_{rel}$, kcal/mol | 4,527769697 | 0 | 1,651874833 | -9,812227999 |
| d(Fe=O), Å | 1.60792 | 1.75673 | 1.74692 | 1.77334 |
| ∠(O-Fe=O) | 98.418<br>108.190 | 124.649<br>120.188 | 113.371<br>125.102 | 119.439<br>118.639 |
| ∠(O-Fe-O) | 108.022<br>134.952 | 115.642<br>125.048 | 115.243<br>125.110 | 114.747<br>125.103 |
| ∠(Fe-O-H) | 178.928 | 114.315 | 123.484 | 135.477 |
| ∠(Fe-O-C) | 179.189 | 104.701 | 120.736 | 134.938 |
| $<S^2>$ | 6.0778 | 6.7541 | 6.7362 | 6.9929 |
| $q_s$(Fe) | 3.143 | 4.046 | 4.017 | 4.093 |
| $q_s$(O) | 0.568 | -0.552 | -0.216 | 0.381 |
| $q_s$(H) | 0.002 | -0.006 | -0.006 | -0.032 |
| $q_s$(C) | -0.002 | -0.007 | -0.291 | -1.053 |
| E(Fe1s), a.u. | -256.212310 | -256.220820 | -256.204890 | -256.200640 |
| E(O1s), a.u. | -19.193480 | -19.194690 | -19.174420 | -19.147700 |
| E(Fe1s), eV. | -6971,89207 | -6972,123639 | -6971,690162 | -6971,574514 |
| E(O1s), eV | -522,2811933 | -522,3141191 | -521,7625443 | -521,0354561 |
| $E_{rel}$(Fe1s), eV. | 0 | -0,231569 | 0,201908 | 0,317556 |
| $E_{rel}$(O1s), eV | 0 | -0,0329258 | 0,518649 | 1,2457372 |

$E_{ads}$ = E(NoSymm+$CH_4$) - (E(NoSymm) + E($CH_4$)) = -6716.72352955-(-6676.19577210-40.5339575449) = 0.006200095 (a.u.) = 3.89061845 (kcal/mol)

*Table 7. The different states of $O=Fe(Ga(OH)_2)_2(\mu-O)_4(Si(OH)_2)_2$ model in $C_{2v}$ point group: energy, geometrical parameters, spin densities, $<S^2>$ and energy of 1s-orbital of Fe and O from group $[FeO]^{2+}$.*

| Symmetry | $^5A_1$ | $^5A_2$ | $^5B_1$ | $^5B_2$ | NoSymm |
|---|---|---|---|---|---|
| Energy, a.u. | -6676.17718700 | | | | -6676.19577210 |
| $E_{rel}$, kcal/mol | | | | | |
| d(Fe=O), Å | 1.60352 | | | | 1.60533 |
| ∠(O-Fe=O) | 117.147 | | | | 111.254<br>100.479 |
| ∠(O-Fe-O) | 125.706<br>125.706 | | | | 135.000<br>113.227 |
| $<S^2>$ | 6.1760 | | | | 6.1103 |
| $q_s$(Fe) | 3.320 | | | | 3.241 |
| $q_s$(O) | 0.256 | | | | 0.430 |

| E(Fe1s), a.u. | -256.210917 | | | -256.209420 |
|---|---|---|---|---|
| E(O1s), a.u. | -19.196999 | | | -19.191630 |
| E(Fe1s), eV. | -6971,854165 | | | -6971,81343 |
| E(O1s), eV | -522,3769502 | | | -522,2308523 |
| $E_{rel}$(Fe1s), eV. | 0 | | | 0,040735 |
| $E_{rel}$(O1s), eV | 0 | | | 0,1460979 |

| Symmetry | NoSymm |
|---|---|
| Energy, a.u. | -4785.17740446 |
| d(Fe=O), Å | 1.60555 |
| ∠(O-Fe=O) | 97.371<br>152.774 |
| ∠(O-Fe-O) | 166.245<br>100.728 |
| <S2> | 6.0603 |
| $q_s$(Fe) | 3.102 |
| $q_s$(O) | 0.736 |
| E(Fe1s), a.u. | -256.218220 |
| E(O1s), a.u. | -19.204390 |
| E(Fe1s), eV. | -6972,05289 |
| E(O1s), eV | -522,5780696 |

$E_{ads}$ = E(NoSymm+CH$_4$) - (E(NoSymm) + E(CH$_4$)) = -4825.71243280-(-4785.17740446-40.5339575449) = -0.001070795 (a.u.) = -0.671934098 (kcal/mol)

*Table 8. Characteristics of adsorbed system, reactants, transition state and products for 6-ring model ZSM-5 (FeAl2Si5H12O21) +CH$_4$*

| Structure | NoSymm+CH$_4$ | Reactants | TS | Products |
|---|---|---|---|---|
| | 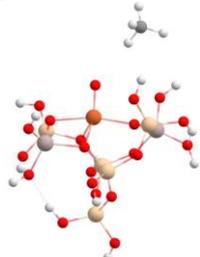 | 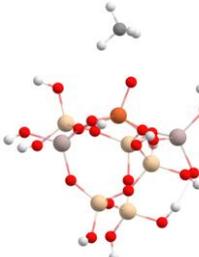 | 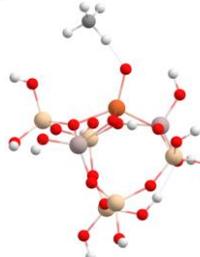 | 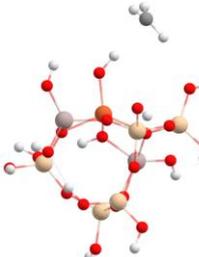 |
| Energy, a.u. | -4825.71243280 | -4825.73380372 | -4825.72808972 | -4825.74702002 |
| $E_{rel}$, kcal/mol | 13,41045532 | 0 | 3,585589283 | -8,293353805 |
| d(Fe=O), Å | 1.60619 | 1.60277 | 1.73064 | 1.78574 |
| ∠(O-Fe=O) | 96.317<br>152.860 | 128.369<br>101.322 | 103.463<br>124.631 | 108.609<br>102.747 |
| ∠(O-Fe-O) | 165.889<br>100.809 | 145.059<br>131.852 | 146.403<br>131.861 | 154.291<br>130.300 |
| ∠(Fe-O-H) | 120.269 | 124.508 | 132.508 | 131.762 |
| ∠(Fe-O-C) | 120.269 | 114.000 | 129.126 | 130.559 |
| <S²> | 6.0607 | 6.1214 | 6.6583 | 6.9896 |
| $q_s$(Fe) | 3.099 | 3.249 | 3.970 | 4.158 |
| $q_s$(O) | 0.733 | 0.425 | -0.167 | 0.413 |
| $q_s$(H) | 0.02 | 0 | -0.008 | -0.031 |
| $q_s$(C) | 0.000 | 0 | -0.205 | -1.069 |
| E(Fe1s), a.u. | -256.218550 | -256.216300 | -256.210160 | -256.203930 |
| E(O1s), a.u. | -19.204070 | -19.202180 | -19.181200 | -19.147940 |
| E(Fe1s), eV. | -6972,06187 | -6972,000644 | -6971,833566 | -6971,664039 |
| E(O1s), eV | -522,5693619 | -522,5179324 | -521,9470375 | -521,0419868 |
| $E_{rel}$(Fe1s), eV. | 0 | 0,061226 | 0,228304 | 0,397831 |
| $E_{rel}$(O1s), eV | 0 | 0,0514295 | 0,6223244 | 1,5273751 |

| Symmetry | NoSymm |
|---|---|
| Energy, a.u. | -8149.80635589 |
| d(Fe=O), Å | 1.60657 |
| ∠(O-Fe=O) | 95.837<br>97.359 |
| ∠(O-Fe-O) | 166.120<br>101.501 |
| <S2> | 6.0595 |
| $q_s$(Fe) | 3.106 |
| $q_s$(O) | 0.734 |
| E(Fe1s), a.u. | -256.214850 |
| E(O1s), a.u. | -19.200190 |
| E(Fe1s), eV. | -6971,961187 |
| E(O1s), eV | -522,4637817 |

$E_{ads}$ = E(NoSymm+CH$_4$) − (E(NoSymm) + E(CH$_4$)) = -8190.34175660 –(-8149.80635589-40.5339575449) = -0,001443165 (a.u.) = -0,90559981 (kcal/mol)

*Table 9. Characteristics of adsorbed system, reactants, transition state and products for **6-ring model ZSM-5 (FeGa2Si5H12O21)** +CH$_4$*

| Structure | NoSymm+CH$_4$ | Reactants | TS | Products |
|---|---|---|---|---|
| | 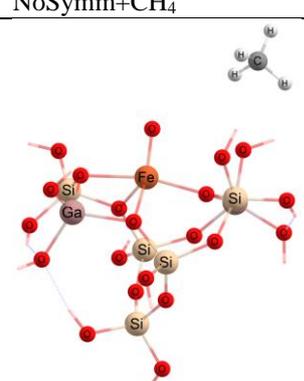 | | 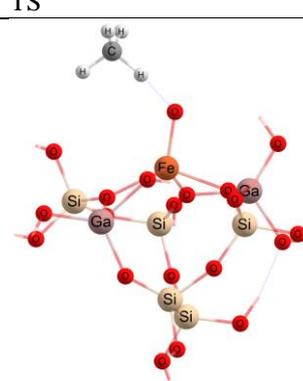 | |
| Energy, a.u. | -8190.34175660 | -8190.36227521 | -8190.35592933 | -8190.37478040 |
| $E_{rel}$, kcal/mol | 12,8756227 | 0 | 3,982099986 | -7,847125525 |
| d(Fe=O), Å | 1.60713 | 1.60357 | 1.74013 | 1.79239 |
| ∠(O-Fe=O) | 95.912<br>153.136 | 123.410<br>104.612 | 123.460<br>110.147 | 95.964<br>101.896 |
| ∠(O-Fe-O) | 165.966<br>101.544 | 154.200<br>134.210 | 147.796<br>133.408 | 146.977<br>134.231 |
| ∠(Fe-O-H) | 122.342 | 147.660 | 130.192 | 128.152 |
| ∠(Fe-O-C) | 117.509 | 142.971 | 126.540 | 122.516 |
| <S²> | 6.0595 | | 6.6823 | |
| $q_s$(Fe) | 3.104 | 3.273 | 4.007 | 4.183 |
| $q_s$(O) | 0.733 | 0.399 | -0.185 | 0.4 |
| $q_s$(H) | 0.001 | 0.002 | -0.005 | -0.017 |
| $q_s$(C) | 0.000 | -0.002 | -0.220 | -1.082 |
| E(Fe1s), a.u. | -256.214910 | | -256.205220 | |
| E(O1s), a.u. | -19.199710 | | -19.174530 | |
| E(Fe1s), eV. | -6971,96282 | | -6971,699142 | |
| E(O1s), eV | -522,4507203 | | -521,7655376 | |
| $E_{rel}$(Fe1s), eV. | 0 | | 0,263678 | |
| $E_{rel}$(O1s), eV | 0 | | 0,6851827 | |